\documentclass[twocolumn,pra,aps,showpacs,nofootinbib]{revtex4-1}

\usepackage{graphicx}
\usepackage{subfigure}
\usepackage{dcolumn}
\usepackage{bbm}
\usepackage{color,epstopdf}
\usepackage{amscd}
\newcommand{\mathd}{\mathrm{d}}

\usepackage{amsmath}

\newcommand{\tmop}[1]{\ensuremath{\operatorname{#1}}}

\begin{document}
\title{Interaction-induced correlations and non-Markovianity of quantum dynamics}
\author{A. Smirne$^1$, L. Mazzola$^2$, M. Paternostro$^2$, and B. Vacchini$^1$}
\address{$^1$Dipartimento di Fisica, Universit\`a degli Studi di Milano, Via Celoria 16, I-20133 Milan, Italy
\& INFN, Sezione di Milano, Via Celoria 16, I-20133 Milan, Italy\\
$^2$Centre for Theoretical Atomic, Molecular and Optical Physics, School of Mathematics and Physics, Queen's University, Belfast BT7 1NN, United Kingdom}
\begin{abstract}
We investigate the conditions under which the trace distance between two different states of a given open system increases in time due to 
the interaction with an environment, 
therefore signalling non-Markovianity. We find that the finite-time difference in trace distance is bounded by two sharply defined quantities 
that are strictly linked to the occurrence of system-environment correlations created throughout their interaction and affecting the subsequent 
evolution of the system. This allows to shed light on the origin of non-Markovian behaviours in quantum dynamics. We best illustrate our findings 
by tackling two physically relevant examples: a non-Markovian dephasing mechanism that has been the focus of a recent experimental endeavour and the 
open-system dynamics experienced by a spin connected to a finite-size quantum spin chain.

\end{abstract}
\maketitle


Recently great attention has been paid to the 
development of a more general understanding of the dynamics of open quantum
systems, in order to deal with the occurrence of memory effects 
\cite{Breuer2002,Budini2004,Piilo2008a,Breuer2008,Shabani2009,Mazzola2009,Chruscinski2010,Barchielli2010,Apollaro2011,Znidaric2011,Vacchini2011,Rosario2012,
Haikka2012,Laine2012,Rosario2012a,Zhang2012,Mazzola2012}. In particular, different definitions
of quantum non-Markovianity have been theoretically introduced \cite{Wolf2008,Breuer2009,Rivas2010,Hou2011}
and, in some cases, experimentally investigated \cite{Liu2011,Tang2012,Chiuri2012,Liu2012}.
Nevertheless, the physical reasons ruling whether an open quantum system exhibits a Markovian
or a non-Markovian dynamics have still to be fully clarified. 

A widely accepted view is that, in Markovian dynamics,
the correlations between the open system and its environment as well as 
the changes in the environmental state due to
the interaction 
do not have a significant influence on the subsequent evolution of the open system.
This picture is often introduced relying on qualitative
considerations, possibly assuming that the total state at time $t$ can be effectively represented
as a product state between the state of the open system
at the time $t$ and a fixed state of the environment \cite{Gardiner2000,Breuer2002}.
It is worth noting how the same assumption also lies at the
foundations of the quantum regression hypothesis \cite{Lax1968,Swain1981,Talkner1985,Ford1996}.
System-environment correlations induced by the interaction and changes in the state of the environment 
are thus thought to be at the basis of non-Markovian dynamics.

In this paper, we show how this relationship can be formulated
in a quantitative way using the properties of the trace distance~\cite{Nielsen2000}, whose time evolution 
can be used to characterise the dynamics of an open quantum system that evolves starting from two different initial states~\cite{Breuer2009,Laine2010b,Breuer2012}.
Any change of the trace distance between states of an open system can be interpreted as 
an exchange of information with the environment that affects it: a non-monotonic behaviour of the trace distance witnesses the fact that some information previously
lost by the open system can affect it back again, thus inducing memory effects in
its evolution. In view of this interpretation, non-Markovian dynamics 
can be identified with those dynamics that show an increase of the trace distance at some intervals of time~\cite{Breuer2009,Breuer2012}.

Here, starting from the analysis reported in Ref.~\cite{Mazzola2012}, we introduce
an upper and a lower bound to the variation of the trace distance at finite time
intervals. The bounds express  quantitatively the influence
of the system-environment correlations and the changes in the state of the environment at a time $t$ on
the subsequent dynamics of the open system. They thus allow to
estimate how system-environment correlations, as well as evolution of
the environment,
account for the Markovian or non-Markovian nature of open system's
dynamics.
In particular our lower bound provides a sufficient condition for the onset of non-Markovianity.
We apply our analysis to two physical examples: the experimental setting considered in Ref.~\cite{Liu2011} that describes the transition from Markovian to
non-Markovian dephasing on a qubit and the energy-non-conserving open-system dynamics of a spin-$1/2$ particles that is coupled to a finite-size quantum spin chain~\cite{Apollaro2011,ApollaroVarie}. 

The remainder of this paper is organized as follows. 
In Sec.~\ref{sec:rse}, we first introduce
an upper and a lower bound to the variation of the trace distance, which hold under very
general conditions. We further show how, as a consequence, 
the strength of the effects of system-environment correlations and environmental evolution can determine the non-Markovian character
of a given dynamics. In Sec.~\ref{sec:ex}, we apply our general analysis to two physically relevant examples. First, we 
address the single-qubit pure-dephasing mechanism exploited in Ref.~\cite{Liu2011}
to investigate experimentally the transition between Markovian and
non-Markovian dynamics. Second, we study the case of a single spin interacting with a finite-size spin environment embodied by a quantum spin chain~\cite{Apollaro2011,ApollaroVarie}. 
Finally Sec.~\ref{sec:ceo} is devoted to conclusions and final remarks.

\section{Role of system-environment correlations and environmental evolution in the dynamics of open quantum systems}\label{sec:rse}

\subsection{Upper and lower bound to the increase of the trace
  distance on a finite time interval}\label{sec:ulb}

We now highlight the relevance of the
correlations between system and environment, as well as of the changes in
the state of the environment, in determining the variation of the
trace distance among system states. As discussed above,
knowledge of the time dependence of this quantity allows to assess the
Markovianity properties of the time evolution, and is therefore of
primary importance. To this aim let us start
from the decomposition of any  joint system-environment state $\rho_{SE}$ as~\cite{Stelmachovic2001}
\begin{equation}\label{eq:rrr}
 \rho_{SE} = \rho_S \otimes \rho_E + \chi_{SE},
\end{equation}
where $\rho_{S}~(\rho_E)$ is the reduced state of the system $S$ (the environment $E$)
and $\chi_{SE}$ (such that $\mbox{tr}_E\left[\chi_{SE}\right] = \mbox{tr}_S\left[\chi_{SE}\right]=0$) 
accounts for the total correlations between the open system and
the environment, as also shown by
the relation $\| \chi_{SE} \|=2 D(\rho_S \otimes \rho_E,\rho_{SE})$.
The total $S$-$E$ system is usually assumed to be closed, so that its evolution
is provided by a one parameter group of unitary operators $\left\{U_t\right\}_{t \geq 0}$, with $t_0=0$ 
initial time. Given the initial total state $\rho_{SE}(0)$, the total state at a time
$t$ is $\rho_{SE}(t) = U_t \rho_{SE}(0) U^{\dag}_t$ and,
as a consequence, the state at time $t+t'$ can be inferred from the state at the time $t$ through
the relation
\begin{equation}\label{eq:utt}
 \rho_{SE}(t+t') = U_{t',t} \rho_{SE}(t) U^{\dag}_{t',t}
\end{equation}
where $U_{t',t}=U_{t+t'}U^{\dag}_t$.

Using Eq.~(\ref{eq:rrr}) twice, we can simply express the difference in
the total states at time $t$ originating from different initial conditions as follows:
\begin{eqnarray}
  \label{eq:diff}
 &&\rho^1_{SE}(t)- \rho^2_{SE}(t)= (\rho^1_S(t)-\rho^2_S(t))\otimes
  \rho^{1}_E(t)\\&&+\rho^{2}_S(t)\otimes (\rho^1_E(t)-\rho^2_E(t))+(\chi^1_{SE}(t)-\chi^2_{SE}(t)).\nonumber
\end{eqnarray}
An equivalent relation is obtained by exchanging
the role of labels $1$ and $2$. The difference between the total states
can thus be split in two contributions, one depending on the
difference between the states of the reduced systems and the other, made up of
the last two contributions at the right hand side (r.h.s.) of Eq.~(\ref{eq:diff}), given
by the comparison between the reduced environmental states and between
the correlations. We thus identify the two quantities
\begin{widetext}
\begin{equation}
F(t',t, \rho^{1,2}) \equiv  D \left( \tmop{tr}_E \left[ U_{t', t}
  (\rho^1_S(t)\otimes \rho^{1}_E(t)) U^{\dag}_{t', t}\right],\tmop{tr}_E \left[ U_{t', t} (\rho^2_S(t) \otimes \rho^{1}_E(t)) U^{\dag}_{t', t}\right]\right)
  \label{eq:f}
\end{equation}
and
\begin{equation}
B(t', t, \rho^{1,2})\equiv \frac{1}{2}\|\tmop{tr}_E \left[ U_{t', t} \left( \rho^{2}_S(t) \otimes (\rho^1_E(t) - \rho^2_E(t))
  \right) U_{t', t}^{\dag} \right] + \tmop{tr}_E \left[
  U_{t', t} \left( \chi^1_{S E}(t) - \chi^2_{S E}(t) \right) U_{t', t}^{\dag}
  \right]\|,\label{chi}
\end{equation}
\end{widetext}
where we have introduced the trace norm of an element $\sigma$ of the set ${\cal T}$ of linear trace class operators as $||\sigma||=\textrm{Tr}[\sqrt{\sigma^\dag\sigma}]$ 
and the trace distance between two statistical operators $\rho^{1,2}$
\begin{equation}
 D(\rho^1, \rho^2) = \frac{1}{2}\left\|\rho^1-\rho^2\right\| = \frac{1}{2}\sum_k |\varrho_k|,
\end{equation}
with $\varrho_k$ the eigenvalues of the self-adjoint traceless operator $\rho^1-\rho^2$ and $0\leq D(\rho^1, \rho^2) \leq 1$. 
$D(\rho^1, \rho^2)$ quantifies the distinguishability~\cite{Nielsen2000} between two given states.
If a system is prepared in one of the two states $\rho^1$ or $\rho^2$ with probability $1/2$ each,
an observer can design an optimal strategy to guess the preparation
with success probability given by $[1+D(\rho^1,\rho^2)]/2$. 

The quantity $F(t',t, \rho^{1,2})$
describes how the distinguishability between the reduced states would
evolve at a time $t+t'$ if the two total states at the time $t$ were
product states, with the same environmental state, thus building on
the first term at the r.h.s. of Eq.~\eqref{eq:diff}.  In a
complementary way, the quantity $B(t', t, \rho^{1,2})$ keeps track of  the effects of correlations and differences in the
environmental states at a time $t$ on the subsequent dynamics of the open system. Thanks
to the contractivity of the trace norm of a self-adjoint operator under the action of any 
completely positive trace-preserving (CPT) linear map \cite{Ruskai1994}, it is straightforward to show that 
$B(t', t, \rho^{1,2})\in[0,2]$. Moreover, it can be null even if the total states at time $t$ are
not product states, so that the correlations, despite being
present, do not have any influence on the evolution of the trace
distance between two reduced states of $S$.

In what follows, we will study the evolution of the distinguishability between couples of reduced states, 
thus describing the information flow between the open system and its environment~\cite{Breuer2009,Laine2010b,Breuer2012}.
The trace distance between two reduced states at time $t$ will be indicated as
\begin{equation}\label{eq:d}
D(t, \rho^{1,2}) \equiv  D(\rho^1_S (t), \rho^2_S (t)),
\end{equation}
where $\rho^j_S(t) = \mbox{tr}_E[U_t
  \rho^j_{SE}(0)U^{\dag}_t]$, $j=1,2$.
Our analysis focuses on the variation of the trace distance at finite time intervals
\begin{eqnarray}
  \Delta D (t', t,  \rho^{1,2}) & \equiv & D(t+t', \rho^{1,2}) - D(t, \rho^{1,2}).  \label{eq:finite}
\end{eqnarray}
Using Eq.~(\ref{eq:utt}) and (\ref{eq:diff}),
the trace distance between two reduced 
states at time $t+t'$ can be expressed as the sum
of two different contributions, which reflect the decomposition
at the r.h.s. of Eq.~(\ref{eq:diff}).
Then, as the trace norm is unitarily invariant
and the partial trace is a CPT map, the contractivity
of the trace norm under CPT maps as well as 
the triangular inequality for trace norm
\begin{equation}
\label{eq:tr}
 \left| \|\sigma\| - \|\sigma'\| \right| \leq \|\sigma - \sigma'\| \leq \|\sigma\| + \|\sigma'\|~~\forall\sigma, \sigma'\in{\cal T},
\end{equation}
 directly lead to the following bounds to the variation of the trace distance on finite
time intervals
\begin{widetext}
\begin{equation}
\label{eq:finite2}
B(t', t, \rho^{1,2}) - F(t',t, \rho^{1,2}) -  D(t, \rho^{1,2})\le  \Delta D (t', t, \rho^{1, 2})  \leq  B(t', t, \rho^{1,2}) +  F(t',t, \rho^{1,2}) -  D(t, \rho^{1,2}).
\end{equation}
\end{widetext}
Eq.~(\ref{eq:finite2}) holds in complete generality, the only requirement being that one takes into account the full
unitary evolution $U_{t}$, which could well be time inhomogeneous.
The upper bound in Eq.~(\ref{eq:finite2}) shows that 
\begin{equation}
B(t', t, \rho^{1,2}) > D(t, \rho^{1,2}) - F(t',t, \rho^{1,2}) \label{cn}
\end{equation}
is a necessary condition to the increase of the trace distance within the time interval $[t,t+t']$. It is important to note
that the r.h.s. of Eq.~(\ref{cn}) is positive as a consequence of the contractivity
of the trace norm. In fact, $F(t',t, \rho^{1,2})$ can be written
as \cite{Mazzola2012}
\begin{equation}
 F(t',t, \rho^{1,2}) = D \left( \Phi_{t', t}\left[\rho^1_S(t)\right],\Phi_{t', t}\left[\rho^2_S(t)\right]\right), \label{eq:f2}
\end{equation}
where 
\begin{equation}
\Phi_{t', t}\left[\rho_S\right] = \tmop{tr}_E \left[ U_{t', t}(\rho_S\otimes \rho^1_E(t)) U^{\dag}_{t', t}\right]
\end{equation}
for any $t$ and $t'$, so that $\Phi_{t', t}$ is a CPT map \cite{Breuer2002}
and $F(t',t, \rho^{1,2}) \leq D(t, \rho^{1,2})$, with $F(0,t, \rho^{1,2}) = D(t, \rho^{1,2})$.
An increase of the distinguishability between reduced states needs the effects
of system-environment correlations to prevail on the contraction
of the trace distance due to the reduced CPT map $\Phi_{t', t}$ obtained
from a total product state at time $t$. Notice that in the limit
$t'\rightarrow 0$ the upper bound in Eq.~(\ref{eq:finite2}) leads to the bound found in Ref.~\cite{Mazzola2012}.
Finally, by applying again the triangular inequality and using 
the contractivity of the trace norm under CPT maps on Eq.~(\ref{eq:finite2}), one gets
the weaker upper bound
\begin{eqnarray}
   &&\Delta D (t', t, \rho^{1, 2})  \leq   D(\rho^{1}_{SE}(t), \rho^1_S(t) \otimes \rho^1_E(t))\\
&& \nonumber +  D(\rho_{SE}^{2}(t), \rho^2_S(t) \otimes \rho^2_E(t)) + D(\rho^1_E(t), \rho^2_E(t)).  \label{eq:finite2b}
\end{eqnarray}
This confirms that an increase of the trace distance in the time interval $[t+t',t]$
calls for system-environment correlations in at least one of the two total states at time $t$, or for different
environmental states 
$\rho_{E}^{1}(t)$ and $\rho_{E}^{2}(t)$. The inequality in Eq.~(\ref{eq:finite2b})
was first derived in Ref.~\cite{Laine2010b} taking $t=0$ as the initial time of the dynamics
and pointing out that an increase of the trace distance above its initial
value witnesses initial correlations or different initial environmental states. 
Indeed, a non-monotonic temporal behavior of the trace distance can be read as an increase
with respect to its value at a previous time.

Up to now we have seen how system-environment correlations
can be in general compatible also with a decrease of the distinguishability between reduced states.
But to what extent is this the case? Can the effects of correlations and environmental evolution 
be arbitrarily strong without inducing any trace-distance increase? This question is answered
by virtue of the lower bound in Eq.~(\ref{eq:finite2}).
A sufficient condition to have an increase of the trace distance in the time interval $[t,t+t']$ is that the effects  
at the time $t+t'$ of the system-environment correlations and environmental states at the time $t$
are strong enough to satisfy
\begin{equation} 
B(t', t, \rho^{1,2}) > D(t, \rho^{1,2}) + F(t',t, \rho^{1,2}). \label{cs}
\end{equation}
Let us emphasize how the fulfillment of this condition 
implies the occurrence, with certainty, of an increase of the trace distance, which is quite a remarkable result.

\subsection{System-environment correlations and quantum non-Markovianity}\label{sec:sos}

In virtue of the bounds to the variation of the trace distance discussed in the previous Subsection,
we can now show that system-environment
correlations and changes in the environmental states actually determine the Markovian/non-Markovian character of a dynamics according to the definition given in Ref.~\cite{Breuer2009}.

Let us assume a product initial state $\rho_{SE}(0)=\rho_S(0) \otimes \rho_E(0)$
with a set environmental state $\rho_E(0)$, which implies the existence
of a well-defined reduced dynamics on the whole set of statistical
operators of the open system. In fact, the open system's dynamics can then be described via a family of CPT maps $\left\{\Lambda(t)\right\}_{t\geq0}$,
with \cite{Breuer2002}
\begin{equation}\label{eq:alpha}
 \Lambda(t)\left[\rho_S\right] = \tmop{tr}_E \left[ U_{t}(\rho_S\otimes \rho^1_E(0)) U^{\dag}_{t}\right], 
\end{equation}
so that $\rho_S(t) = \Lambda(t) \rho_S(0)$.
Non-Markovianity of quantum dynamics has been defined and quantified in terms of the various properties
of this family of CPT maps. In particular, 
the measure of non-Markovianity $\mathcal{N}(\Lambda)$ introduced in~\cite{Breuer2009}
can be expressed as
\begin{equation}\label{eq:blp}
 \mathcal{N}(\Lambda) =  \max_{\rho^{1, 2}} \sum_k \Delta D (b_k-a_k, a_k, \rho^{1, 2}),
\end{equation}
where $(a_k, b_k)$ are the time intervals where the trace distance $D(t, \rho^{1,2})$ increases,
and the maximum over all pairs of initial reduced states is taken.
${\cal N}(\Lambda)$ quantifies the total amount of information
that flows to the open system, as witnessed by the trace distance,
and  the relevance of memory effects on the reduced dynamics.
From Eq.~(\ref{eq:blp}), it is clear that a quantum dynamics
is non-Markovian if and only if there is a time interval $[t,t+t']$
and a pair of initial states $\rho^{1,2}_{S}(0)$ of the system such that $\Delta D (t', t, \rho^{1, 2})>0$.
As we have assumed an initial total product state, 
with fixed environmental state, the dependence on $\rho^{1, 2}$ should be intended from now on as
referred to the two initial reduced states. In addition, 
possible system-environment correlations in the total states,
as well as differences between the environmental states, at a time $t$
have entirely to be ascribed to the system-environment interaction up to time $t$.

\begin{figure}
\begin{center}
\includegraphics[width=\columnwidth]{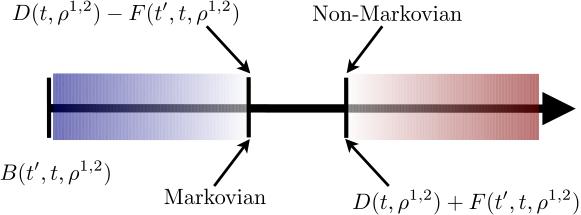}
\caption{{\small{The horizontal axis represents the possible values of $B$ defined in Eq.~(\ref{chi}) at fixed times
$t$ and $t'$ and initial reduced states $\rho^1_S(0)$ and $\rho^2_S(0)$, while the two marks
represent the thresholds given by  $D(t, \rho^{1,2})-F(t',t, \rho^{1,2})$ (blue region, lower mark) and 
$ D(t, \rho^{1,2}) + F(t',t, \rho^{1,2})$ (red region, upper mark), with 
$D(t, \rho^{1,2})$ and $F(t',t, \rho^{1,2})$ defined in Eq.~(\ref{eq:d}) and Eq.~(\ref{eq:f}) respectively.
If $B(t',t, \rho^{1,2})$ lies above the upper threshold
\emph{for some} $t, t', \rho_S^{1,2}(0)$, the dynamics is non-Markovian, while if it takes values  below the 
lower threshold \emph{for all} the $t, t', \rho_S^{1,2}(0)$, the dynamics is Markovian. 
Values of  $B(t',t, \rho^{1,2})$ between the two thresholds are compatible with both Markovian and non-Markovian dynamics. }}}
\label{fig:1}
\end{center}
\end{figure}

Eqs.~(\ref{cn}) and (\ref{cs}) 
provide a general reference scale that relates the Markovian or non-Markovian
nature of a given dynamics to 
the relevance of the quantity $B(t', t, \rho^{1,2})$
with respect to $D(t, \rho^{1,2})$ and $F(t',t, \rho^{1,2})$. This
relation is schematically depicted in Fig.~\ref{fig:1}, where the horizontal axis
represents the possible values of $B(t', t, \rho^{1,2})$ together with
the two
reference values given by $ D(t, \rho^{1,2}) \pm F(t',t, \rho^{1,2}) $. The line above (below) the upper (lower) threshold denotes the region where non-Markovianity (Markovianity) is enforced. For the intermediate region
of width $2F(t',t, \rho^{1,2})$ no general statement can be made
from the two bounds. Indeed, even if system-environment correlations and evolution of the environment due to the interaction affect the dynamics of the open system, this does not guarantee that the dynamics is non-Markovian.
But the reduced dynamics is surely non-Markovian if the effects of system-environment correlations 
and environmental evolution, as quantified by $B(t', t, \rho^{1,2})$, 
are strong enough to exceed the upper threshold in Fig.~\ref{fig:1}.

In addition, $B(t', t, \rho^{1,2})$ gives an indication of the degree of non-Markovianity of the dynamics, as expressed by
the measure $\mathcal{N}(\Lambda)$ in Eq.~(\ref{eq:blp}). Any variation of the trace distance
on a finite time interval can be lower bounded through Eq.~(\ref{eq:finite2}), so that
one could introduce a lower bound to $\mathcal{N}(\Lambda)$. This is
larger the more the effects of system-environment correlations and environmental evolution
rise above the upper threshold in Fig.~\ref{fig:1}.

As a further remark, let us notice that our analysis relates the (non-)Markovianity of a given evolution
with the general correlations between the system and its environment, regardless of their (quantum or classical) nature. The relevant quantity $B(t', t, \rho^{1,2})$
is defined in terms of the total correlations $\chi_{SE}^j(t)$, which include
both classical and quantum correlations. In general, the interaction does not have to build up
quantum correlations in order to determine a reduced non-Markovian dynamics~\cite{Pernice2011}.

\section{Examples}\label{sec:ex}
\subsection{Transition from Markovian to non-Markovian dephasing}
\label{uno}

We now consider, as our first explicit example, the model described in Ref.~\cite{Liu2011},
where the transition from Markovian to non-Markovian dephasing dynamics has been 
experimentally realized by modifying the initial state of the environment.
The total system under investigation consists of single photons generated
by spontaneous parametric down conversion and passing through a Fabry-P{\'e}rot cavity
mounted on a rotator and then through a quartz plate.
The open-system qubit is encoded in the space spanned by two orthogonal polarization states, while the frequency of the photons
is used to encode the environment. 
The dephasing dynamics experienced by the polarization of light and due to the quartz plate can be described through the unitary system-environment evolution
\begin{equation}\label{eq:u}
U(t) | \lambda, \omega \rangle = e^{i n_{\lambda} \omega t} | \lambda, \omega \rangle \quad (\lambda = H, V),
\end{equation}
where $|{H}\rangle, |{V}\rangle$ denote horizontal and vertical polarization states, $n_{\lambda}$ is the refractive index of the plate for a $\lambda$-polarized incident photon, and $|\omega\rangle$ stands for the environmental state at set frequency $\omega$.
The initial state of the environment is $\rho_E(0) = |\psi_E(0)\rangle \langle \psi_E(0)|$ with
\begin{equation}\label{eq:ine}
|\psi_E(0)\rangle = \int \mathd \omega f(\omega) |\omega \rangle,
\end{equation}
and the frequency distribution $|f(\omega)|^2$ is controlled 
via the tilting angle of the Fabry-P{\'e}rot cavity.
For any fixed amplitude $f(\omega)$, the dynamics of polarization 
can be described by a family of CPT maps $\left\{\Lambda(t)\right\}_{t\geq0}$ such that 
\begin{equation}\label{eq:masd}
 \Lambda(t) \left(\begin{array}{cc}
    \rho_{HH} & \rho_{HV} \\
    \rho_{VH}& \rho_{VV} 
  \end{array}\right) = \left(\begin{array}{cc}
    \rho_{HH} & k^*(t) \rho_{HV} \\
   k(t) \rho_{VH}& \rho_{VV} 
  \end{array}\right), 
\end{equation}
where $\rho_{\lambda \lambda'}=\langle\lambda|\rho_S(0)|\lambda'\rangle~(\lambda,\lambda'=H,V)$ with $\rho_S(0)$ the initial polarization state, 
and the time-dependent dephasing function
\begin{equation}\label{eq:k}
k(t) = \int \mathd \omega |f(\omega)|^2 e^{i (n_V-n_H) \omega t}.
\end{equation}
has been introduced.
Regardless of the choice of $f(\omega)$, the pair of initial system states
that maximizes the increase of trace distance 
is
\begin{equation}
\label{eq:ins}
|\psi^\pm_S(0)\rangle=\frac{1}{\sqrt{2}} \left( |H\rangle \pm |V\rangle \right).
\end{equation}
The calculation of the corresponding trace distance at a time $t$ is straightforward and leads us to  
 $\mathcal{N}(\Lambda) = \sum'_j (|k(b_j)|-|k(a_j)|)$,
where the sum is taken over the temporal region of extremes $a_j$ and $b_j\ge a_j$ where $|k(t)|$ grows.
The dynamics at hand is thus non-Markovian iff
\begin{equation}\label{eq:cns}
|k(t+t')| > |k(t)|
\end{equation}
for some $t, t' \geq 0$.
By adjusting the distribution $f(\omega)$, one can arrange for a transition from 
Markovian to non-Markovian open-system dynamics~\cite{Liu2011}, which can be
characterized through the time-local master equation
\begin{equation}\label{eq:tcl}
\frac{\mathd}{\mathd t} \rho_S(t) = -i \epsilon (t) \left[\sigma_z , \rho_S(t)\right] + \gamma(t) \left(\sigma_z \rho_S(t) \sigma_z - \rho_S(t)\right),
\end{equation}
where
$\epsilon(t) = -\frac{1}{2} \mbox{Im}[\partial_t\ln k(t)]$ and 
$\gamma(t) = -\frac{1}{2} \mbox{Re}[\partial_t\ln k(t)]$.
\begin{figure*}[!ht]
{\bf (a)}\hskip5cm{\bf (b)}\hskip5cm{\bf (c)}\\
\includegraphics[width=0.7 \columnwidth]{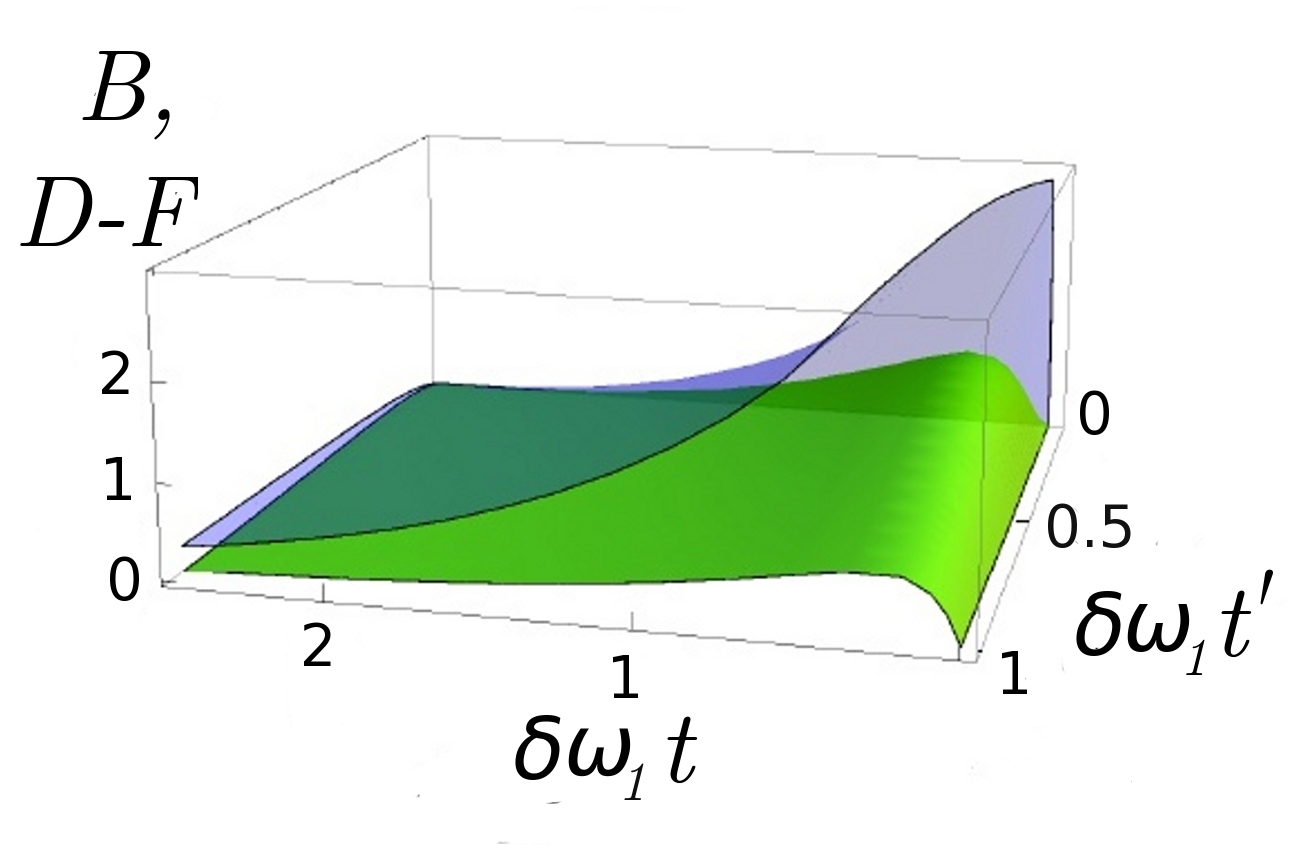}\includegraphics[width=.7\columnwidth]{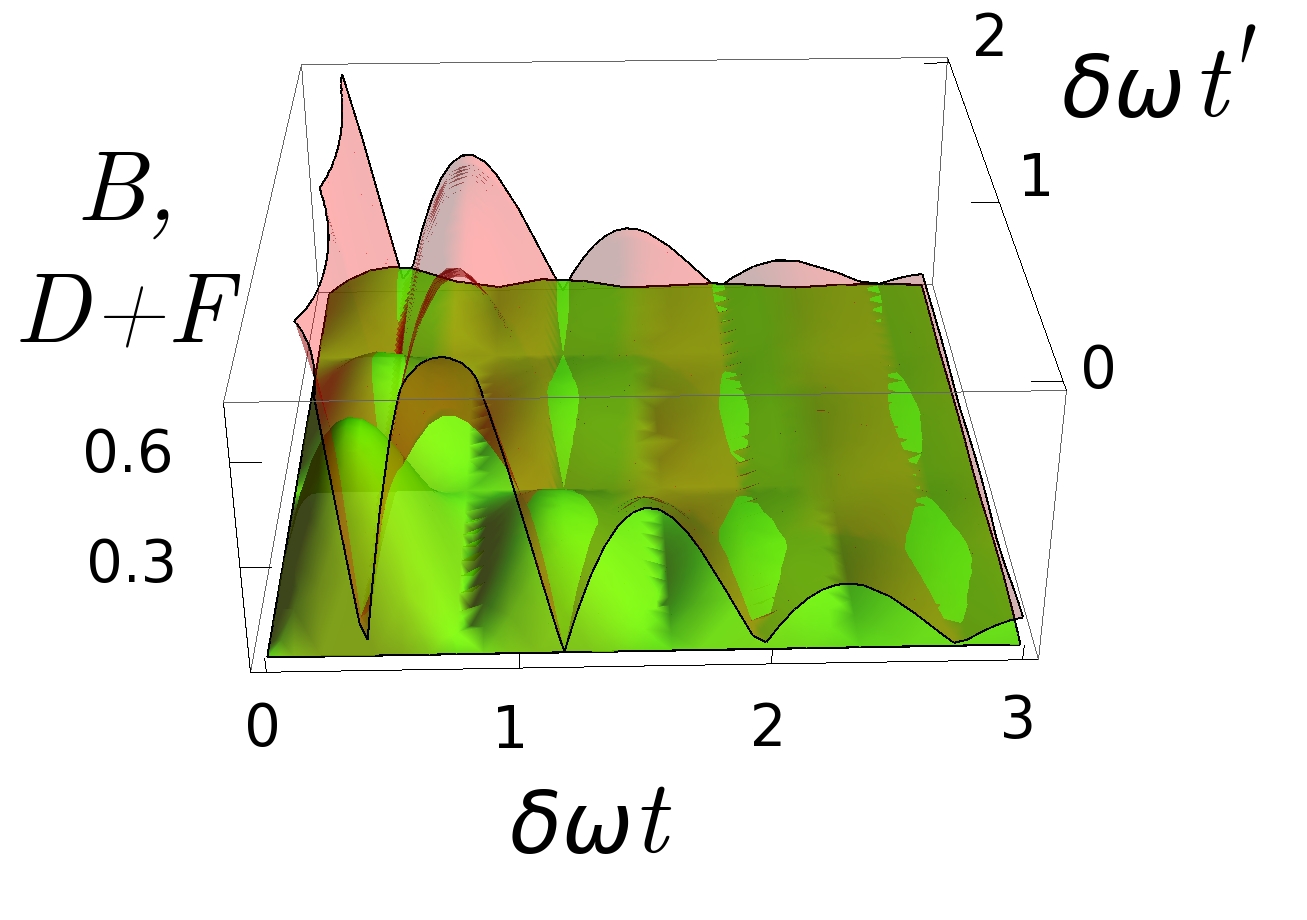}\includegraphics[width=.7\columnwidth]{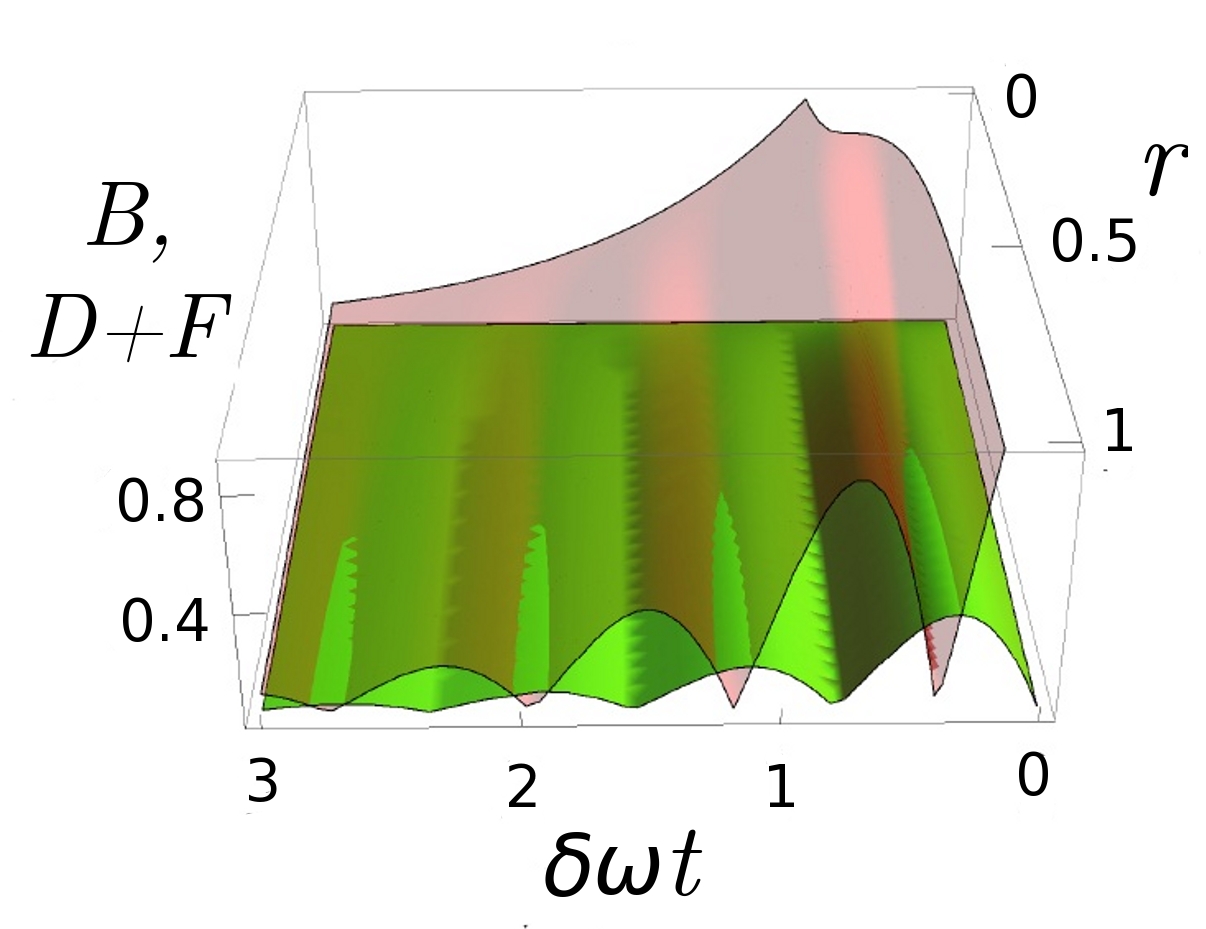}
\caption{Plot of the effects of system-environment correlations
described through the function $B(t',t'\rho^{1,2})$ given in Eq.~(\ref{eq:es1}) corresponding to the example discussed in Sec.~\ref{uno} for a
frequency distribution as in Eq.~(\ref{eq:dlor}). 
Panel {\bf (a)}: we have taken $\omega^0_1/\delta\omega_1 =
\omega^0_2 / \delta\omega_1$, $\delta\omega_2 / \delta\omega_1= 10$, and $r=1$. In both panel {\bf (b)} and {\bf (c)} we have taken
$\delta\omega_{1,2}\equiv\delta\omega$, $\omega^0_1/\delta\omega = 1$ and $\omega^0_2/\delta\omega=9$ with $r=1$ [$t' \delta\omega=0.3$] 
in panel {\bf (b)} [panel {\bf (c)}]. Semi-transparent surfaces represent $D(t,\rho^{1,2})-F(t',t,\rho^{1,2})$ [panel {\bf (a)}]
and $D(t,\rho^{1,2})+F(t',t,\rho^{1,2})$ [panels {\bf (b)} and {\bf (c)}].}
\label{fig:2}
\end{figure*}

Let us now address how our general analysis applies to this model. The transition
from Markovian to non-Markovian dynamics is shown by following the
evolution of the trace distance for the pair of initial reduced
states in Eq.~(\ref{eq:ins}) upon variation of the initial state of the
environment, as set by $f(\omega)$.
Thus, we can focus on the evolution of 
the initial total states $|\psi^1_{SE}(0)\rangle = |\psi^+_S(0)\rangle\otimes |\psi_E(0)\rangle$
and  $|\psi^2_{SE}(0)\rangle = |\psi^-_S(0)\rangle\otimes |\psi_E(0)\rangle$. 
The corresponding total
states at time $t$  
are found using Eq.~(\ref{eq:u}). In turn, this allows us to evaluate analytically all the quantities of interest introduced
in Sec. \ref{sec:rse}, which are defined in terms of the trace norm of operators on the Hilbert space of the open system. 
It turns out that the two environmental states are equal at all times, 
so that the possible increases of the trace distance can be traced back solely to the correlations built during the interaction between the open system
and the environment. 
In particular, 
we find that the effects of 
system-environment correlations at time $t$ on the subsequent
evolution of the open system are quantified through
\begin{equation}\label{eq:es1}
 B(t', t, \rho^{1,2}) = \left| k(t+t') - k(t) k(t') \right|.
\end{equation}
Moreover, we have that 
$F(t', t, \rho^{1,2})  =   \left| k(t) k(t') \right|$ and 
 $\Delta D (t', t, \rho^{1, 2}) = |k(t+t')|-|k(t)|$.
Therefore, Eq.~(\ref{eq:finite2}) is indeed satisfied. 
These quantities can be explicitly evaluated by specifying the
frequency distribution $|f(\omega)|^2$ that determines $k(t)$ through Eq.~(\ref{eq:k}).
As a first example, we use a Lorentzian distribution: as we shall see, this leads to a semigroup evolution and
thus provides a natural benchmark. Explicitly, we take
\begin{equation}\label{eq:ldis}
|f(\omega)|^2=\frac{\delta\omega}{\pi[(\omega-\omega^0)^2 +
  \delta\omega^2]},
\end{equation}
where $\omega^0$ is the central frequency and $\delta\omega$ is the width
of the Lorentzian distribution, which can be obtained using single photons emitted by quantum dots~\cite{Tang2012}.
Choosing a Lorentzian frequency distribution corresponds to taking 
 $k(t) = e^{(i \omega_0 - \delta \omega) t}$, 
which entails the exponential decay of the trace distance $D(t, \rho^{1,2}) = e^{- \delta \omega t} $.
In this case, the coefficients of the time-local generator
in Eq.~(\ref{eq:tcl}) reduce to $\epsilon(t) = \omega_0/2$ and $\gamma(t) = \delta \omega/2$, so that the dynamics of the open system is fixed
by a completely positive semigroup~\cite{Gorini1976, Lindblad1976}.
In addition, the exponential expression of $k(t)$ means
that the correlations between the open system and the environment have no influence on the evolution of the
trace distance between reduced states, as it follows from Eq.~(\ref{eq:es1}).
Thus, for the model at hand, in the semigroup regime system-environment correlations, despite being present,
do not affect the dynamics of the open system at all.
The exponential expression for $k(t)$ leads to $B(t', t,
\rho^{1,2}) = 0$, which in turn implies 
$  D (t' + t, \rho^{1, 2})=  F(t', t, \rho^{1,2})$: States $\rho^j_{SE}(t)$ $(j=1,2)$
can be replaced with $\rho^j_S(t) \otimes \rho_E(t)$ without modifying the subsequent evolution 
of the trace distance.

Let us now consider the frequency distribution 
\begin{equation}\label{eq:dlor}
 |f(\omega)|^2= \sum_{j=1,2}\frac{A_j}{(\omega-\omega^0_j)^2 + \delta\omega^2_j},
\end{equation}
so that
$k(t) = [e^{(i \omega^0_1- \delta\omega_1)t} + r e^{(i \omega^0_2- \delta\omega_2)t}]/({1+r})$,
with $r = A_1/A_2$. As it can be easily checked, if the two Lorentzian distributions in Eq.~(\ref{eq:dlor}) have
the same central frequencies, $\omega^0_1 = \omega^0_2$, the resulting dynamics
is still Markovian as $|k(t)|$ is monotonically decreasing. Incidentally, differently from the case of a single Lorentzian distribution, the family of
CPT maps determined by Eq.~(\ref{eq:masd}) is no longer a semigroup but a divisible family of completely positive dynamical maps~\cite{Rivas2010}. In fact, 
the coefficients of the corresponding generator in Eq.~(\ref{eq:tcl})
are given by
\begin{equation}\label{coeff2}
\epsilon(t) = \frac{\omega_0}{2},  \qquad \gamma(t) = \frac{\delta \omega_1 e^{-\delta \omega_1 t} + r \delta \omega_2 e^{-\delta \omega_2 t} }{2(e^{-\delta \omega_1 t}+r e^{-\delta \omega_2 t} )},
\end{equation}
which are positive at all times. Our analysis allows 
to trace the Markovianity of the dynamics back to the weakness of the system-environment correlations 
created by the interaction. 
\begin{figure*}[!ht]
{\bf (a)}\hskip6cm{\bf (b)}\hskip6cm{\bf (c)}\\
\includegraphics[width=.6\columnwidth]{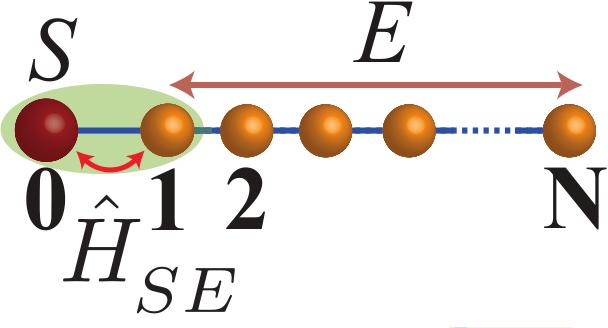}\includegraphics[width=1.5\columnwidth]{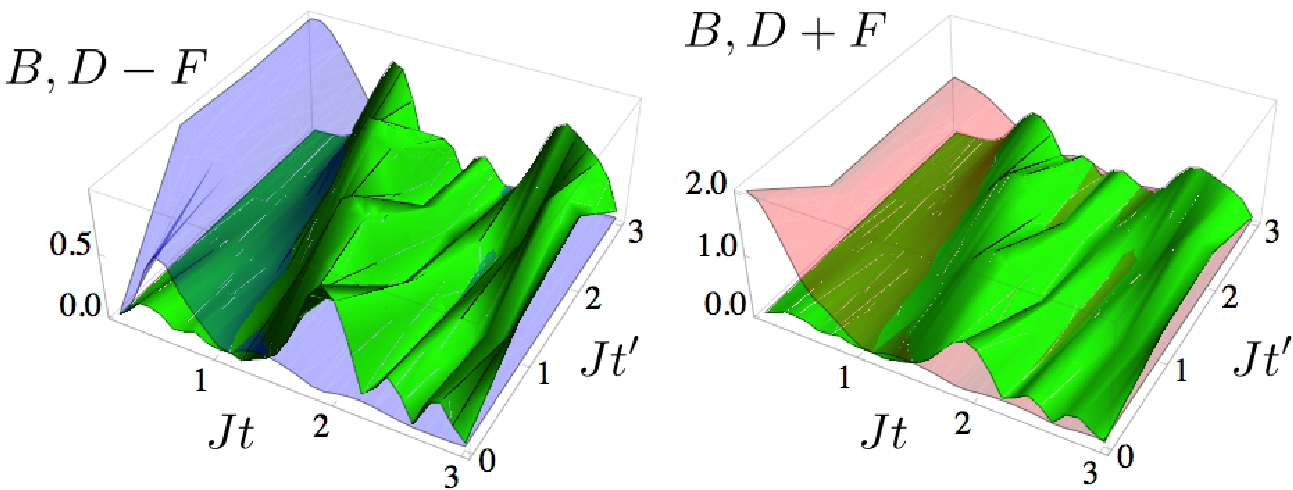}
\caption{{\bf (a)} We consider a quantum spin chain of $N+1$ particles partitioned into a two-level {system} spin embodied by particle $0$ and a 
finite-size environment (provided by the rest of the chain). The dynamics experienced by spin $0$ can be changed from Markovian to strongly 
non-Markovian by adjusting the parameters entering the system-environment coupling model $\hat H_{SE}$ and the inter-environment one $\hat H_E$. 
Details of the form of such Hamiltonians are given in the body of the paper.  {\bf (b)} [{\bf (c)}] Comparison between the 
function $B(t',t,\rho^{1,2})$ (full-colorer surface) and the threshold for Markovianity [non-Markovianity] 
$D(t,\rho^{1,2})-F(t',t,\rho^{1,2})$ [$D(t,\rho^{1,2})+ F(t',t,\rho^{1,2})$]. (semi-transparent curves). 
We have used $N=8$ with $J_0/J=1$ and $B/J=10^{-2}$.}
\label{fig:3}
\end{figure*}
In Fig.~\ref{fig:2} {\bf (a)}, we plot $B(t', t, \rho^{1,2})$ and $D(t, \rho^{1,2})-F(t',t, \rho^{1,2})$.
Both the quantities refer to the two initial states of the system in Eq.~(\ref{eq:ins}) and are plotted
as a function of $t$ and $t'$. The effects of system-environment correlations
on the reduced system, as quantified through $B(t', t, \rho^{1,2})$, are always weaker than the threshold embodied by 
$D(t, \rho^{1,2})-F(t',t, \rho^{1,2})$. As a consequence, the distinguishability between the states of the open system cannot 
increase and a Markovian dynamics is induced. 

A different situation occurs if the two Lorentzian distributions in Eq.~(\ref{eq:dlor}) have the
same width but different central
frequencies. In this case, the dynamics is non-Markovian, i.e. $|k(t)|$ is a non-monotonic function
of time. As shown in Fig.~\ref{fig:2}  {\bf (b)}, the effects of
system-environment correlations are now stronger than the upper
threshold $D(t, \rho^{1,2})+ F(t',t, \rho^{1,2})$.
Basically, there are times $t$ such that the two reduced states, $\rho^1_S(t)$ and $\rho^2_S(t)$ are very similar,
so that the upper threshold $D(t, \rho^{1,2})+ F(t',t, \rho^{1,2})$ is small, while
the correlations of the two total states, $\chi^1_{SE}(t)$ and $\chi^2_{SE}(t)$, are still different. Consequently, 
the trace distance at subsequent times $t+t'$ increase.
In this case, system-environment correlations do induce a non-Markovian dynamics.
Finally, by taking $r$ in Eq.~(\ref{eq:dlor}) from zero to any non-zero value, we
have a transition from a Markovian dynamics, more precisely a semigroup dynamics, to a non-Markovian dynamics. In Fig.~\ref{fig:2} {\bf (c)}
one can see how this is reflected in the behavior of $B(t', t, \rho^{1,2})$.
In fact, the latter turns from being identically zero for $r=0$ to increasing above the upper
threshold $D(t, \rho^{1,2})+ F(t',t, \rho^{1,2})$, thus implying a non-Markovian dynamics.

\subsection{Non-Markovianity in a spin-chain system}

As a second instance, we consider the case of a single qubit attached to a quantum spin chain of $N$ spin-$1/2$ particles, along the 
lines of the studies reported in Ref.~\cite{ApollaroVarie,Apollaro2011} and as illustrated in Fig.~\ref{fig:3} {\bf (a)}. 
The open system is embodied by spin $0$, while the environment consists of particles $1\to{N}$. The overall $N+1$ system are mutually 
coupled via an XX model and subjected to a transverse magnetic field. Assuming units such that $\hbar=1$, 
the corresponding Hamiltonian model is $\hat H{=}\hat H_{SE}+\hat H_E$ with
\begin{equation}\begin{aligned}\label{XX}
& \hat H_{SE}=-2J_0(\hat\sigma_0^x \hat\sigma_{1}^x+ \hat\sigma_0^y \hat\sigma_{1}^y),\\
& \hat H_E=-2J\sum_{n=1}^{N-1}(\hat\sigma_n^x \hat\sigma_{n+1}^x+ \hat\sigma_n^y \hat\sigma_{n+1}^y)-2B\sum_{n=1}^{N} \hat\sigma_n^z,
\end{aligned}\end{equation}
where $\hat\sigma^{k}_n$ is the $k$-Pauli matrix ($k{=}x,y,z$) for particle $n$, $B$ is the amplitude of the magnetic field 
affecting $S$ and $J$ ($J_0$) is the inter-environment (system-environment) coupling strength. The underlying assumption is that the free evolutions 
of $S$ and $E$ are identical, thus allowing the passage to the interaction picture without the introduction of time-dependent coefficients. 
The non-Markovian evolution experienced by spin $0$ was characterized fully in Ref.~\cite{Apollaro2011}, where it was found that, 
for interaction times that are within the recurrence time of the system (when any information propagating across the chain returns to the open system 
after reaching the end of the chain), there is a working point defined by $(J_0/J,B/J)$ at which the measure of non-Markovianity ${\cal N}$ is null. 
As the optimization inherent in the definition of such measure is achieved for system states lying on the equatorial plane of the 
Bloch sphere~\cite{Apollaro2011}, we consider the input states $\rho_S^{\pm}(0)=|\psi^\pm_S(0)\rangle\langle\psi^\pm_S(0)|$, while 
the environment is initialized in $\rho^E_{1,2}(0)=\rho^E_{ini}=\otimes_{i=1}^{N-1}|{0}\rangle_i\langle{0}|$. 
In order to provide a physically significant example that is nevertheless able to 
show clearly the features that we are interested in, we solved fully the problem embodied by an environment 
of $N=8$ spins with $J_0/J=1$ and $B/J=10^{-2}$. At variance with the previous example, the system lacks of an analytically amenable solution 
(the expressions for the trace distance and the quantities introduced in the Section~\ref{sec:rse} are too involved to be reported here) 
but allows for a handy numerical analysis. The results are shown in Fig.~\ref{fig:3} {\bf (b)} and {\bf (c)}, where $B(t,t',\rho^{1,2})$ is compared 
to the thresholds $D(t,\rho^{1,2})\pm F(t,t',\rho^{1,2})$ within a broad range of values for $t$ and $t'$. Clearly, besides the existence of a range 
of values of $(t,t')$ where the dynamics is expectedly Markovian, $B(t,t',\rho^{1,2})$ soon trespasses the thresholds for 
non-Markovianity [cf. Fig.~\ref{fig:3} {\bf (c)}]. In particular, this is the case for values of $(t,t')$ such that $J\cdot(t+t')<3$, which guarantee 
that the corresponding evolutions occur well within the recurrence times of the spin chain and any non-Markovian effect is due to the intrinsic 
features of the interaction rather than the finiteness of the environment. Interestingly enough, the gap between the lower and upper threshold 
identified above may disappear, in this example. 
Values of $t$ exist at which $D(t,\rho^{1,2})\simeq F(t,t',\rho^{1,2})$, thus making the gap between upper and lower threshold effectively null.   

\section{Conclusions}\label{sec:ceo}

We have shown how the dynamical correlations
between an open quantum system and its environment influence the nature of the open system's dynamics, as far as non-Markovianity is concerned.
Our analysis relies on the definition of non-Markovianity given in terms of the evolution
of the trace distance between reduced states~\cite{Breuer2009}.

We have introduced the quantity $F(t',t, \rho^{1,2})$ [cf. Eq.~(\ref{eq:f})] that describes the evolution of the
trace distance under the CPT maps that would connect states at different
times if correlations and environmental changes could be neglected at
any time. In a complementary way, we have defined
the quantity $B(t', t, \rho^{1,2})$ [see Eq.~(\ref{chi})] that
measures, by means of the trace norm, the
effects of system-environment correlations and environmental evolution due to the interaction
up to a time $t$ on the subsequent dynamics of the open system.
These two quantities allow to introduce an upper and a lower bound to the variation 
of the trace distance between reduced states on finite time
intervals, as quantified by Eq.~(\ref{eq:finite2}). We have thus been able to
conclude that if the effects of correlations and environmental evolution are below a 
first threshold the resulting reduced dynamics is certainly Markovian. 
Despite being necessary, system-environment correlations and changes
in the environment are
not a priori sufficient to induce an increase of the trace distance. On the other hand, 
if their effects exceed a second threshold, a non-Markovian reduced dynamics is surely induced.

The general analysis has been applied to the model exploited in \cite{Liu2011}
to experimentally detect the transition between Markovian and non-Markovian dynamics. We have shown
how such transition can be explained in terms of the different effects
of system-environment correlations. By properly varying
the initial state of the environment, one can describe both a semigroup dynamics,
in which system-environment correlations do not affect at all the evolution of the reduced state's distinguishability,
and non-Markovian dynamics, in which system-environment correlations strongly influence
the dynamics of the open system.

Our results will be useful also with respect to different approaches to non-Markovianity relying on other properties
of the dynamical maps, since they show in full generality to what extent system-environment
correlations can be compatible with a contraction of the trace distance. 
In particular, this could help to further understand the connection between correlations in the total state 
and breaking of divisibility of the completely positive dynamical maps~\cite{Shabani2009,Rosario2012}. 
In addition, our results could
provide further insights into microscopic derivations of reduced dynamics, in order to clarify
the role of system-environment correlations and changes in the environmental state~\cite{Possanner2012}.

\acknowledgments
 AS and BV acknowledge financial support from COST Action MP1006. LM is supported by the EU through 
 a Marie Curie IEF Fellowship. MP thanks the UK EPSRC for a Career Acceleration Fellowship and a grant of the ÒNew Directions for Research LeadersÓ initiative (EP/G004579/1).

\end{document}